\documentclass[prb,notitlepage,showpacs,twocolumn]{revtex4-1}

\usepackage{graphicx,float, amsmath}
\usepackage[dvipsnames]{xcolor}
\usepackage{natbib}
\usepackage{mathrsfs}
\usepackage{epstopdf}
\usepackage{subfigure} 
\usepackage[integrals]{wasysym} 
\usepackage{nicefrac,amssymb,amsmath} 
\usepackage{changes} 

\begin{document}
 
\title{Excitations in a spin--polarized two--dimensional electron gas}
\author{Dominik Kreil}
\author{Raphael Hobbiger}
\author{J\"urgen T. Drachta}
\author{Helga M. B\"ohm}
\affiliation{Institut f\"ur Theoretische Physik, Johannes Kepler Universit\"at, 4040 Linz, Austria}
\begin{abstract}
A remarkably long--lived spin plasmon may exist in two--dimensional electron liquids with imbalanced spin up and spin down population.
Predictions for this interesting mode by \citeauthor{agarwal2014long} [Phys.\ Rev.\ B {\bf90},\ 155409 (2014)] are based on the random phase approximation.
We here show how to account for spin dependent correlations from known ground state pair correlation functions and study the consequences on the various spin dependent longitudinal response functions.
The spin plasmon dispersion relation and its critical wave vector for Landau damping by minority spins turn out to be significantly lower.
We further demonstrate that spin dependent effective interactions imply a rich structure in the excitation spectrum of the partially spin--polarized system.
Most notably, we find a ``magnetic antiresonance'', where the imaginary part of both, the spin--spin as well as the density--spin response function vanish.
The resulting minimum in the double--differential cross section is awaiting experimental confirmation.

\end{abstract}
\maketitle

\section{Introduction}

In non--magnetic electron layers, ({\it i.e.\/}\ with  vani\-shing spin--polarization 
\(P\!\equiv (N_\uparrow\!-\!N_\downarrow)/N\), total number of particles 
\(N\!= N_\uparrow\!+\!N_\downarrow\), and \(N_\sigma\) being the number of electrons with spin 
up or down), collective spin modes rapidly decay into electron--hole pairs.
For spin--polarized systems, however, as was convincingly demonstrated by \citeauthor{agarwal2014long}\cite{agarwal2014long}, 
the Random Phase Approximation (RPA) yields an amazingly long--lived spin plasmon.
This ``longitudinal magnon'' exists, {\it inside\/} the band of electron--hole pairs with the majority spin, up to a critical wave vector \(q_{\rm spl}^{\rm max}\) before decaying rapidly into electron--hole pairs of the minority spin population.

In this report we show, among other results, that going beyond the traditional RPA leads to much lowered critical wave vectors. 
For both, the conventional plasmon as well as the spin plasmon, this effect becomes more pronounced for dilute systems, where correlations play an important role.

Experiments\cite{nagaoplasmon,hirjibehedin2002evidence} on heterostructures were performed on the conventional plasmon ({\it i.e.}\ the \(P\!=\!0\) charge plasmon) for areal densities \(n\!=1.9\!\times\! 10^{13}\,\mathrm{cm}^{-2}\) and \(n\!=(0.77\ldots4) \times\! 10^{9}\,\mathrm{cm}^{-2}\).
This corresponds to  \emph{Wigner--Seitz} radii \(r_\mathrm{s}\!\lessapprox\!2\) and 
\(r_\mathrm{s}\!\approx 10\ldots20\), respectively 
(as usual, \(r_{\mathrm{s}}\!\equiv 1/(a_{\scriptscriptstyle\rm B}^*\sqrt{\pi n}\)) 
with \(a_{\scriptscriptstyle\rm B}^*\) being the material's effective Bohr radius). 
The influence of electron correlations on the dispersion can be estimated using
the simulation based\cite{moroni1992staticA} charge--charge response function
from Ref.\ \onlinecite{davoudi2001analytical}.
At the coupling parameters of interest, as shown in Table \ref{TAB:vphresults}, the critical wave vector 
\(q_{\mathrm{pl}}^{\mathrm{max}}\) for Landau damping changes by typically \(\ge\!\)\;20\% .
Only in systems of sufficiently low \(r_\mathrm{s}\) and for low wave vectors \(q\) the mode can be detected near the RPA result.
Possible candidates for such dense systems could be electron gases near a SrTiO\(_3\) surface\cite{meevasana2011creation, santander2011two, maryenko2012temperature, wang2014anisotropic, HaoDiebold2015Coexistence} with a high background dielectric constant \(\epsilon_{\scriptscriptstyle\mathrm b}\); 
(however, anisotropy effects need to be accounted for, too\cite{Gokmen2007Spin}).

For the spin plasmon, we here show that spin dependent correlations pull this mode down drastically towards the minority particle--hole band.
Consequently, it appears questionable whether this excitation can be resolved experimentally, even if it may stay slightly above the boundary.
We also give results for low densities and predict a new phenomenon.

In Sec.~\ref{sec: Theory} we briefly describe the theory. 
We first introduce our method to account for spin dependent correlations via effective (static) interactions\cite{gori2004pair}, and then study the consequences for the various response functions.
In Sec.~\ref{sec: Results} we critically discuss our results.
Special emphasis is put on the critical wave vector for Landau damping of the
charge plasmon and a detailed investigation of the spin plasmon.
Finally, we present a hitherto unknown valley in the imaginary part of the longitudinal spin 
response, the ``magnetic antiresonance'', and summarize our conclusions in Sec.~\ref{sec: Conclusion}.

\section{Theory}
\label{sec: Theory}

\subsection{Spin dependent effective interactions}
\label{ssec: Spin--dependent effective interactions}

In RPA--type approaches the partial response functions \(\chi_{\sigma\sigma'}\) forming the matrix \(\boldsymbol\chi\) are determined by the equation
\cite{giuliani2005quantum}
\begin{equation}
 \boldsymbol\chi^{-1}(q,\omega) =\>
 \mbox{\(\boldsymbol\chi^0\)}^{-1}(q,\omega) -
  \mathbf{V}(q) \quad.
 \label{eq: matrixeqRPA}
\end{equation}
Here, \(\boldsymbol\chi^0\) contains the spin--resolved parts \(\delta_{\sigma\sigma'}\, \chi^{0}_{\sigma}\) of Stern's polarizability\cite{stern1967polarizability},
and \({ \mathbf{V} }\) the effective interactions \(V_{\sigma\sigma'}\)
between electrons of spin \(\sigma\) and \(\sigma'\). (Eq.\,(\ref{eq:
matrixeqRPA}) may also be read as the definition 
\footnote{
It remains questionable, however, whether this way of packing all 
 non-mean-field effects into dynamic effective potentials optimally
 elucidates the relevant physics.
}
of dynamic interactions \({ \mathbf{V}(q,\omega) }\)).
In the bare RPA studied by Agarwal {\it et al.\/}\cite{agarwal2014long} all
\(V_{\sigma\sigma'}\) are replaced with the Coulomb interaction, 
\( v(q)= 2\pi e^2/(\epsilon_{\scriptscriptstyle\mathrm b} q) \). 

For a paramagnetic layer, {\it i.e.}\ \(P\!=\!0\), various static approximations
have been presented~\cite{[{See, e.g., the references in Ref.\
\protect{\onlinecite{giuliani2005quantum}}; and in }] [{}]
Reinholz2012Dielectric, [{}] [{}] tanatareurophys1997,
NSSS91, sarmaPhysRevB.64.165409, [{}] [{ (for bulk systems)}] Barriga2009Dynamical}.
Commonly, the effective spin dependent interactions are expressed via 
so--called ,,local field corrections'',
\begin{equation}
 V_{\sigma\sigma'}(q) =\> v(q)\,
 \big(1- G_{\sigma\sigma'}(q)\,\big) \;.
 \label{eq: VqGq}
\end{equation}
We term approaches of type (\ref{eq: VqGq}) ``Generalized RPA'' (GRPA). 

The matrix equation (1) for {\(\boldsymbol\chi\)} reads explicitly ({\it c.f.\/}
 Eq.\,(1) of Ref.~\onlinecite{agarwal2014long})
\begin{equation}
 \begin{pmatrix} \chi_{\uparrow\uparrow} & \chi_{\uparrow\downarrow} \\ 
     \chi_{\uparrow\downarrow} &\chi_{\downarrow\downarrow} \end{pmatrix}^{-1}
 =\>
 \begin{pmatrix} \chi^0_{\uparrow} & 0 \\ 0
    &\chi^0_{\downarrow} \end{pmatrix}^{-1} -\>
 \begin{pmatrix}
    V_{\uparrow\uparrow} & V_{\uparrow\downarrow} \\
    V_{\uparrow\downarrow} & V_{\downarrow\downarrow} \end{pmatrix} \;\;,
 \label{eq: matrixeqRPA explicit}
\end{equation}
where we also invoked the symmetry \((\uparrow\downarrow)\,\longleftrightarrow\,(\downarrow\uparrow)\).

As pointed out by E.\ Krotscheck\cite{KroTrieste}, an essential requirement for
a response function is to fulfill the first and zeroth moment sum rule. The latter invokes the spin--resolved static structure factors
\begin{equation}
 S_{\sigma\sigma'}(q) \equiv \frac1{\sqrt{N_{\sigma^{\phantom{,\!}}} N_{\sigma'}}}\,
   \langle\delta\widehat n _{\bf q\sigma^{\phantom{,}} }
   \delta\widehat n _{-\bf q\sigma'} \rangle \;\;,
 \label{eq: Ssp def}
\end{equation}
with the partial density fluctuation operator \(\delta\widehat n _{\bf q\sigma}\)
and the prefactor convention of Gori-Giorgi et al.\cite{gori2004pair}.
Again, for non--interacting fermions, \({\mathbf S^0}\), the matrix of static structure factors, is diagonal\cite{giuliani2005quantum}.
The full static structure factor is given by \(S(q) =
\sum_{\sigma\sigma'}S_{\sigma\sigma'\!}(q)\,\sqrt{n_\sigma n_{\sigma'}}/n\).

The pertinent sum rules then read
\begin{subequations}
 \begin{align}
 &-\int\limits_{0}^\infty\!\frac{d\omega}{\pi}\>
     \mathrm{Im}\,\chi_{\sigma\sigma'}(q,\omega)  \>=\>
     \sqrt{n_\sigma n_{\sigma'}}\> S_{\sigma\sigma'}(q) \;,
  \label{subeq: M0 SR}
 \\
 &-\int\limits_{0}^\infty\!\frac{d\omega}{\pi}\>
     \omega\, \mathrm{Im}\,\chi_{\sigma\sigma'}(q,\omega)  \>=\>
  \delta_{\sigma\sigma'}n_\sigma\,\frac{\hbar\,q^2}{m} \;,
  \label{subeq: M1 SR}
 \end{align}
 \label{eq: SRs}%
\end{subequations}
(\(m\) being the effective electron mass due to the semiconductor background
lattice).
\goodbreak

In order to determine \(V_{\sigma\sigma'}(q)\) from these conditions we replace, as a first 
step, \(\chi^0_{\sigma\sigma'}\) in Eq.~\ref{eq: SRs} with a single--pole (also called 
``collective'') approximation\cite{*[{A similar approach was followed for \(P\!=\!0\) by }]
[{ using \(V_{\uparrow\uparrow}\!\pm\!V_{\uparrow\downarrow}\).}] Asgari2006Static}.
This allows us to derive a compact expression relating the effective interactions with the spatial structure. 
Introducing the matrix \(\bar{\mathbf V}\) of spin weighted interactions via \(\bar V_{\!\sigma\sigma'}(q) \equiv \> \textstyle
  \sqrt{n_\sigma n_\sigma'}/n\displaystyle \, V_{\!\sigma\sigma'}(q)\),
we arrive at the matrix equation
\begin{equation}
 {\bar{\mathbf V}}(q) \>=\> 
 \frac{\hbar^2q^2}{4mn}\, \Big({\mathbf S}^{-2}(q) - {\mathbf S^0}^{-2}(q)\Big) \ .
 \label{eq: VPH_MATRIX_FORM}
\end{equation}
Result (\ref{eq: VPH_MATRIX_FORM}) is the analogue of the particle--hole potential\cite{KroTrieste} defined as
\begin{equation}
 \bar V_{\mathrm{ph}}(q) \>=\> 
 \frac{\hbar^2q^2}{4mn}\, \Big( \frac1{S(q)^2} - \frac1{{S^0(q)}^2} \Big) \;.
 \label{eq: Vphdef}
\end{equation}
The strength of this formula is to contain an approximate summation of both, ladder-- and ring--diagrams, 
thus capturing important long-- as well as short--ranged attributes\cite{polish}. 
Spelling out Eq.~(\ref{eq: VPH_MATRIX_FORM}) explicitly, we obtain
\begin{subequations}
 \begin{align}
  D(q) \;=&\;\; S_{\uparrow\uparrow}(q)\;S_{\downarrow\downarrow}(q) - S_{\uparrow\downarrow}^2(q)
  \phantom{\Big|_|} \;, 
 \\
 \bar V_{\uparrow\downarrow}(q) \>=& -\frac{\hbar^2q^2}{4mn}\, \frac{S_{\uparrow\downarrow}(q) \,
  \big[S_{\uparrow\uparrow}(q) + S_{\downarrow\downarrow}(q)\big]}{D^2(q)} \;,
 \\
 \bar V_{\uparrow\uparrow}(q) \>=& \phantom{-|}\frac{\hbar^2q^2}{4mn}\, \bigg[
  \frac{S_{\downarrow\downarrow}^2(q)+S_{\uparrow\downarrow}^2(q)}{D^2(q)} - \frac1{{S^0_\uparrow}^2(q)} \bigg] \;,
 \end{align}
 \label{eq: no gentleman}%
\end{subequations}
and the analogous expression for \(\bar V_{\downarrow\downarrow}\). 
These interactions can now be used in Eq.~(\ref{eq: matrixeqRPA explicit})
to calculate the response functions from any given set of spin--resolved static
structure factors \(S_{\sigma\sigma'}(q)\).

Note that we do {\it not\/} calculate the response functions --- neither the
spin plasmon nor any other feature --- within the above plasmon--magnon--pole approximation. 
The latter only served the purpose of obtaining suitable effective spin dependent interactions. 
As discussed in Ref.\ \onlinecite{polish}, Eq.~(\ref{eq: Vphdef}) can be seen as the {\it definition\/} of an optimal static effective interaction if the ground state structure factor is known.

High quality spin--resolved ground state structure calculations were
performed by Gori-Giorgi et al.\cite{gori2004pair}. With
reptation quantum Monte Carlo (QMC) techniques they obtained the 
pair--distribution functions \(g_{\sigma\sigma'}(r)\).
A Fourier transform yields the static structure factors we need:
\begin{equation}
 S_{\sigma\sigma'}(q) \>=\> 
 \delta_{\sigma\sigma'} + \sqrt{n_\sigma n_{\sigma'}}\int d^2r\> 
 \big[ g_{\sigma\sigma'}(r) \!-\!1\big]\,e^{i\bf q\cdot r} \,.
 \label{eq: Ssp from gsp} 
\end{equation}
Naturally, all QMC data are limited in real space. Hence an extension \(g_{\sigma\sigma'}(r\!\to\!\infty)\) is necessary in order to establish the proper long--wavelength behavior.
Using reduced units \(\bar q\!\equiv q/k_{\mathrm{F}}\) where \(k_{\mathrm{F}}\!=\!\sqrt{2\pi n}\), and denoting spins opposite to \(\sigma\) as \(\bar\sigma\), this limit reads\cite{gori2004pair} 
\begin{eqnarray}
  S_{\sigma\sigma'}(\bar{q}\to 0) &=& \tilde\xi_{\sigma\sigma'}\,\frac{\bar q}{\pi} +
  \frac{\sqrt{n_{\sigma}n_{\sigma'}}}{n}\,\frac{\bar{q}^{3/2}}{2^{3/4}\,\sqrt{r_\mathrm{S}}} +
  {\cal O}(\bar{q}^2)
 \nonumber \;, \\ 
  \tilde\xi_{\sigma\sigma'} &=&
  \delta_{\sigma\sigma'}\,\sqrt{n_{\bar{\sigma}}/n_\sigma} \,- 
  \delta_{\sigma\bar{\sigma'}} .
  \label{eq: Sssp qto0}
\end{eqnarray}
For the spin--summed \(g(r)\) at any \(P\), as well as for the partial \(g_{\sigma\sigma'}(r)\) 
at \(P\!=\!0\) and \(P\!=\!1\), analytical expressions are given in Ref.\ \onlinecite{gori2004pair}.
These are based on skillful extrapo\-lation to large \(r\) and we follow this procedure for the \(P\!=\!0.48\) data\footnote{Raw data Monte-carlo data provided by P.
Gori-Giorgi.}.
The delicate behavior of the effective interaction between minority spins, \(V_{\downarrow\downarrow}(q)\), 
necessitates additional care with respect to ensuring the high--density (RPA) limit of the fit
for all partial \(S_{\sigma\sigma'}(q)\).

With these results for \({\mathbf S}\) the effective interactions \({\mathbf V}\) are now obtained from (\ref{eq: VPH_MATRIX_FORM}).
This is then used in the matrix equation (\ref{eq: matrixeqRPA explicit}) to determine 
\(\boldsymbol\chi\).

As a check, we evaluated the sum rules (\ref{eq: SRs}) for the spin--summed charge--charge response function. 
The f--sum rule (\ref{subeq: M1 SR}) is excellently fulfilled for all \(r_\mathrm{s}\), the input \(S(q)\) is reproduced within a few percent of error.

\subsection{Response functions}
\label{ssec: Resonse functions}

In an electron liquid subject to an electrostatic external potential \(V^{\mathrm{ext}}\) and an uniaxial 
magnetic field \(\mathbf{B}^{\mathrm{ext}}\) the induced partial spin densities
\(\delta n_\sigma\) manifest themselves in the following observables: the induced particle density 
\(\delta n\!= \delta n_\uparrow \!+\! \delta n_\downarrow\), 
the induced longitudinal magnetization proportional to \(\delta s\!\equiv \delta n_\uparrow \!-\! 
\delta n_\downarrow\), and transverse magnetization components. 
The Pauli spin--flip operators
govern the transverse linear response functions; 
their eigenmodes are the ''conventional'' magnons of condensed matter
physics.
Longitudinal excitations are fully decoupled\cite{polinitosiMB06,
[{This is also fortunate for spin density functional theory }] [{}] Eich2013Transverse}.
Rescaling the magnetic field by Bohr's magneton and the \(g-\)factor, 
\(b^\mathrm{ext} \equiv g\mu_{\scriptscriptstyle\mathrm{B}}|{\bf B}|^\mathrm{ext}/2 \),
we have
\begin{equation}
 \begin{pmatrix} \delta n \\ \delta s \end{pmatrix} \>=\>
 \begin{pmatrix} \chi_{nn} & \chi_{ns} \\ 
                 \chi_{ns} & \chi_{ss} \end{pmatrix} \cdot
 \begin{pmatrix} V^\mathrm{ext} \\ b^\mathrm{ext} \end{pmatrix} \;.
 \label{eq: longitud response}
\end{equation}
Apparently, in electron layers with imbalanced spin popu\-lation a longitudinal
magnetization can be induced either by a magnetic field or by an electrostatic 
potential (or both).
The corresponding longitudinal magnon, {\it i.e.\ } the collective mode in the
spin density \(\,\delta s\,\) at vanishing \(b^\mathrm{ext}\) and \(V^\mathrm{ext}\) is 
termed\cite{agarwal2014long} ''spin plasmon''.

Decomposed into their spin--resolved contributions the relevant susceptibilities 
read
\begin{subequations}
 \begin{align}
 \chi_{nn} &= \chi_{\uparrow\uparrow} + 2\;\chi_{\uparrow\downarrow} +
\chi_{\downarrow\downarrow} \label{eq: chi_nn} \;,\\
 \chi_{ss} &= \chi_{\uparrow\uparrow} - 2\;\chi_{\uparrow\downarrow} +
\chi_{\downarrow\downarrow} \label{eq: chi_ss} \;,\\
 \chi_{ns} &= \chi_{\uparrow\uparrow} - \chi_{\downarrow\downarrow}   \;.           
 \label{eq: chi_ns}
 \end{align}
 \label{eq: all susceptibilities}
\end{subequations}
They all share the same denominator \(\Delta\), 
\begin{equation}
 \Delta =  1 - V_{\uparrow\uparrow}\chi^0_\uparrow
    -  V_{\downarrow\downarrow}\chi^0_\downarrow
    + \big(V_{\uparrow\uparrow}V_{\downarrow\downarrow} - 
           V_{\uparrow\downarrow}^2\big)\,\chi^0_\uparrow\chi^0_\downarrow \;.\quad
 \label{eq: denominator}
\end{equation}

For arbitrary spin--polarization $P$ all \(V_{\sigma\sigma'}\) differ. In the paramagnetic case 
the symmetry $V_{\uparrow\uparrow} \!=\! V_{\downarrow\downarrow}$ implies that
\begin{equation}
 \Delta^{\!P=0} =  1 - V_{\uparrow\uparrow}\,(\chi^0_\uparrow + \chi^0_\downarrow)
    + \big(V_{\uparrow\uparrow}^2 - V_{\uparrow\downarrow}^2\big)\,\chi^0_\uparrow\chi^0_\downarrow \;.
 \label{eq: denominator zeta0}
\end{equation}
For spin independent interactions, as in the bare RPA, Eq.\ (\ref{eq: denominator})
reduces to
\begin{equation}
 \Delta^{\scriptscriptstyle\mathrm{RPA}} \,= 1 - v\,(\chi^0_\uparrow + \chi^0_\downarrow)
 =\, \epsilon^{\scriptscriptstyle\mathrm{RPA}} \ .
 \label{eq: denominator RPA}
\end{equation}
Obviously, the effective interactions change the collective excitations compared to their RPA behavior due to two distinct causes:
The difference between like and unlike spins arising from the Pauli principle influences systems with arbitrary \(P\): from Eq.~(\ref{eq: denominator zeta0}) it is seen that a denominator of the type
\(1\!- V\chi^0\) with {\it some\/} interaction \(V(q)\) would require
\(V_{\uparrow\uparrow}\!=\!V_{\uparrow\downarrow}\,\), also for \(P\!=\!0\). 
In addition, the substantially different screening between minority and majority
components, manifest in \(V_{\uparrow\uparrow}\!\ne\!V_{\downarrow\downarrow}\,\),
gives rise to further modifications for spin--imbalanced systems.

For completeness, we also list the numerators involved in Eq.~(\ref{eq: all susceptibilities}),
\begin{subequations}
 \begin{align}
 \chi_{{\scriptstyle nn}\atop{\scriptstyle ss}} &= \frac1\Delta\,\Big(
 \chi^0-\big[V_{\uparrow\uparrow}\!+\!V_{\downarrow\downarrow}\mp\!2V_{\uparrow\downarrow}\big]
  \chi^0_{\uparrow}\chi^0_{\downarrow}\Big)  \;,\\
 \chi_{ns} &= \frac1\Delta\,\Big( \chi^0_{\uparrow}\!-\!\chi^0_{\downarrow} +
  \big[V_{\uparrow\uparrow}\!-\!V_{\downarrow\downarrow}\big]\,\chi^0_{\uparrow}\chi^0_{\downarrow}\Big) \;. 
 \end{align}
 \label{eq: all susc numerators}
\end{subequations}
We now turn to the numerical results of our approach.

\section{Results}
\label{sec: Results}

\subsection{Charge plasmon}
\label{ssec: charge plasmon}

We start with studying the critical wave vector \(q^{\mathrm{max}}_{\mathrm{pl}}\) of the \(P\!=\!0\) charge plasmon.
This is insensitive to whether using Eq.\ (\ref{eq: VPH_MATRIX_FORM}) or (\ref{eq: Vphdef}).
Table \ref{TAB:vphresults} compares our data with those following from Ref.\ \onlinecite{davoudi2001analytical} based on simulations\cite{moroni1992staticA} for \(\omega\!=\!0\).
\begin{table}[h]
\begin{tabular}{c|cccccc}
\(r_\mathrm{S_{\phantom|}}\)     &    2    &    5    &    10    &    20  &    30    &    40 \\
\(n_{\scriptscriptstyle\mathrm{GaAs}}\) \(\left[10^{9}\,\mathrm{cm}^{-2}\right]\)    
                                 &    75.2& 12    & 3    &  0.75 & 0.33  &  0.19 \\
\hline
& \multicolumn{6}{c}{\(q^{\mathrm{max}}_{\mathrm{pl}}\)} \\ \hline
RPA    \(\left[k_\mathrm{F}^{-1}\right]\)        & 1.50 & 2.45 & 3.55 & 5.09 & 6.28 & 7.29 \\
RPA \(\left[10^{5}\,\mathrm{cm}^{-1}\right]\) &    10.3 & 6.75 & 4.88 & 3.50 & 2.88 & 2.51 \\
\hline
& \multicolumn{6}{c}{change from RPA\(^{\phantom{\big|\!}}\)} \\ \hline
GRPA - Ref.\,\onlinecite{davoudi2001analytical}  & -25\% & -40\% & -52\% & -     & -     & -\\
GRPA - Eq.\,(\ref{eq: Vphdef})  & -15\% & -37\% & -50\% & -62\% & -68\% & -71\%  \\
\end{tabular}
\caption{Paramagnetic charge plasmon critical wave vector.
Upper two lines: (bare) RPA value in reduced units and for a GaAs quantum well.
Lower two lines: Percental change due to the local field corrections of Davoudi
{\it et al.\/}\cite{davoudi2001analytical} and with \(G(q)= 1\!-
V_{\mathrm{ph}}(q)/v(q)\) based on Monte Carlo \(S(q)\) from Ref.\,\onlinecite{gori2004pair}.
}
\label{TAB:vphresults}
\end{table} 
Considering that, by contrast, \(V_\mathrm{ph}(q)\) arises from an \(\omega-\)integration, it is 
striking how close the values are for \(r_\mathrm{s}\!\approx5\ldots10\,\).
The discrepancy at small \(r_\mathrm{s}\) is removed if \(V_\mathrm{ph}(q)\) is determined numerically 
from the sum rule (\ref{subeq: M0 SR}) without the single mode approximation.
This strongly supports the quality of our approach.
In the following we prefer to stick to the analytic relations (\ref{eq: no gentleman}) between the effective 
interactions and the static structure factors, in favor of better physical
insight.

\subsection{Spin plasmon}
\label{ssec: Spin plasmon}

\begin{figure}[htb]
 \centering
  \includegraphics[width=0.5\textwidth]{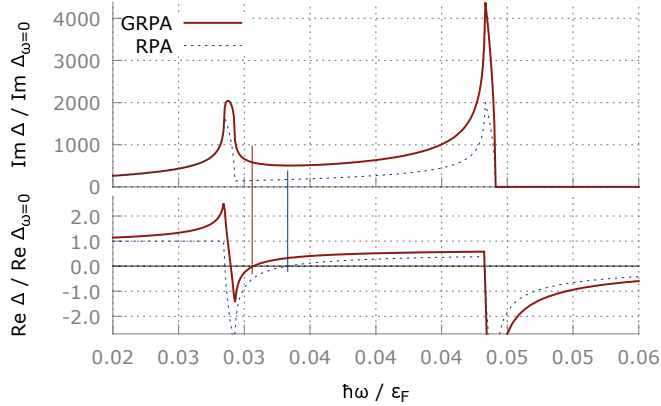}
  \caption{
   Denominator of the spin response functions \(\chi_{\sigma\sigma'}(q,\omega)\) for \(r_\mathrm{s}\!=\!2\), 
   \(q \!= 0.02\,k_\mathrm{F}\), and polarization \(P\!=\!0.48\approx\!0.5 \) in bare RPA (dotted lines) 
   and our GRPA (full curves).
   The upper (lower) panel gives the imaginary (real) part; the vertical lines mark the zeroes of the real part. 
   Terms \(\propto\!\chi^0_\uparrow\chi^0_\downarrow\) as in  Eq.\,(\ref{eq:
denominator zeta0}) enter both, \(\Delta(q,\omega)\) as well as the numerators
of the \(\chi_{\sigma\sigma'}\), changing the overall height of both. 
For better comparability, we thus rescale the curves.
}
\label{FIG:EXAMPLE_CUT}
\end{figure}

Concerning both, charge-- and spin response, Fig.~\ref{FIG:EXAMPLE_CUT} compares the real and imaginary part of the denominator \(\Delta(q,\omega)\) of the susceptibilities (\ref{eq: all susceptibilities}) with its RPA counterpart, \(\epsilon^{\scriptscriptstyle\mathrm{RPA}}(q,\omega)\).
The same system parameters and wave vector are chosen as in  Fig.\ 2a of Ref.\ \onlinecite{agarwal2014long}.
In the GRPA the typical ``shark--fin'' structure of the imaginary part is smoothened for the
minority band and enhanced for the majority spins. 
Like in bare RPA, also for spin--sensitive effective interactions the real part of the denominator has an additional zero above the first band edge.
This zero was identified in Ref.\ \onlinecite{agarwal2014long} as the spin plasmon and in careful investigations proven to be quite stable against 
damping by impurity scattering.

\begin{figure}[H]
 \centering
 \includegraphics[width=0.5\textwidth]{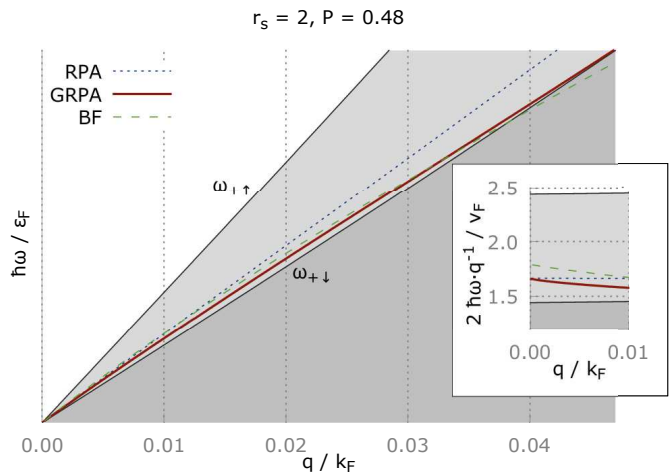}
 \caption{
  Spin plasmon dispersion for \(r_\mathrm{s}\!=\!2\) and \(P\!=\!0.48\)
  in bare RPA (blue dotted line), GRPA (with the effective interactions of 
  Eq.\,(\ref{eq: VPH_MATRIX_FORM}), red solid line) and  single--pole
  approximation (green dashed line).
  All three modes start in the conti\-nu\-um of the majority spins (light 
  grey area). Our result enters the minority spin continuum (dark grey area)
  tangentially at a much lower 
  \(q_{\mathrm{spl}}^{\mathrm{max}}/k_{\scriptscriptstyle\mathrm{F}}\) than 
  that of the RPA.
  The inset shows the dispersion divided by the wave vector, demonstrating that 
  our approach gives the same initial slope as the RPA.
\label{FIG:SPIN_PLASMON_DISPERSION}}
\end{figure}

The spin plasmon, if a true collective mode and pole of \(\chi_{ss}\)\,, can be obtained from either of the two equivalent requirements:
\begin{equation}
  \mathrm{Re}\,\Delta(q,\omega) = 0 
  \;\;\Leftrightarrow\;\; 
  -\mathrm{Im}\, \chi_{ss}(q,\omega) = \mathrm{max} \;\;.
  \label{eq: zero of Re denominator}
\end{equation}
Inside the particle--hole band of the minority spins the two routes do not yield exactly the same result.
We follow Ref.\ \onlinecite{agarwal2014long} by determining the dispersion from the roots of \(\mathrm{Re}\,\Delta(q,\omega)\).
Fig.~\ref{FIG:SPIN_PLASMON_DISPERSION} shows the numerically obtained zeros for \(r_\mathrm{s}\!=\!2\) and \(P\!=\!0.48\) in the \((q,\omega)-\)plane. 
For comparison, the RPA and the single--mode result (``Bijl-Feynman type'' or ``BF'')\cite{giuliani2005quantum} are 
displayed as well.
The inset of Fig.\ \ref{FIG:SPIN_PLASMON_DISPERSION} confirms that our spin dependent GRPA recovers the high density ({\it i.e.}\ RPA) limit. 

It is seen that the inclusion of spin effects in the effective potentials \(V_{\sigma\sigma'}\) lowers the spin plasmon's position significantly. 
As the zero of \(\,\mathrm{Re}\,\Delta(q,\omega)\) is shifted towards lower frequencies, it is simultaneously moved closer to the ``fin structure'' which is smeared out by spin--correlation effects
({\it cf.\/} Fig.~\ref{FIG:EXAMPLE_CUT}, upper part).
In addition, the relative height of \(\,\mathrm{Im}\,\Delta(q,\omega)\) is
larger in the GRPA, implying that damping\citep{fetter2003quantum} of the mode
is stronger everywhere.
Both effects, the close vicinity to Landau damping by minority spins
as well as the overall increase of \(\mathrm{Im}\,\Delta(q,\omega)\) 
heighten the challenge for experimentally verifying the position of this mode.

\begin{figure}[H]
 \centering
  \includegraphics[width=0.5\textwidth]{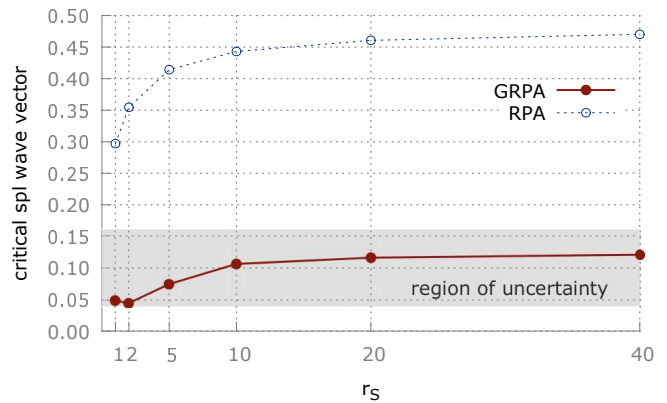}
  \caption{
   Critical wave vector \(q_{\rm spl}^{\rm max}(P\!=\!0.48)\) in our GRPA (red solid line) 
   and bare RPA (blue dotted line) versus coupling strength.
   The shaded area is an estimate of the error induced by the limited \(r-\)range of the 
   input Monte Carlo data\cite{gori2004pair}. 
 \label{FIG:QC_SPIN_PLASMON}}
\end{figure}

Since the spin plasmon, being an acoustic mode, is rather close to the relevant band edge for all \(q\), its critical wave vector for Landau damping is much smaller than that of the charge plasmon. Consequently, 
while the effective interactions \(V_{\sigma\sigma'}(q)\) appear rather un\-affected by minor variations in \(g_{\sigma\sigma'}(r\!\to\!\infty)\), the spin plasmon is quite sensitive to such changes. 
Reducing these uncertainties would require the exact \(q^{>3/2}\) expansion coefficients of \(S_{\sigma\sigma'}(q\!\rightarrow\! 0)\). 
In Fig.~\ref{FIG:QC_SPIN_PLASMON} we present our results for the critical wave vector \(q_{\rm spl}^{\rm max}\), where the spin plasmon tangentially hits the band edge \(\omega_{+\downarrow}\).

It is evident that exchange--correlation effects lower \(q_{\rm spl}^{\rm max}\) to approximately one third of its RPA value.
Even if we account for a substantial spread in the \(q^{5/2}\) coefficient of \(S_{\sigma\sigma'}(q\!\rightarrow\! 0)\), the
reduction is still 50\%.
In order to reduce the uncertainty in the \(r\!\to\!\infty\) input data high accuracy calculations of \(S(q)\) in this regime are desirable (e.g.\ via the so--called ``FHNC'' method\cite{egger2011bose}).
Both, the RPA and the GRPA yield a nearly density independent critical wave vector beyond
\(r_{\mathrm{s}}\!\gtrsim\!10\), as it is typical for static effective
interactions. Investigations in the dynamic many body approach\cite{bohm2010dynamic}
are under way. This holds the promise of a ``charge plasmon revival''\ \cite{godfrin2012observation} at large wave vectors, as first observed in the pioneering work of Neilson {\it et al.\/}\ \cite{NSSS91}.

\subsection{Fermi Liquid paramters}
\label{ssec: Fermi Liquid paramters}

At the densities of interest the Coulomb energy is at least of similar order as the kinetic energy. 
Nevertheless the Fermi surface is remarkably robust, as captured by Landau's Fermi liquid theory
\cite{polinitosiMB06}\(^,\)
\cite{*[{Helpful introductions are also found in }][{ and }]
         ColemanMBP2015, *SolyomSolids32011}.
Low lying excitations behave like quasiparticles; their 
Fourier expanded interaction defines the Landau parameters (rescaled
with the (true) density of states at the Fermi surface)\cite{Iwam91,KwCM94}
\begin{eqnarray}
 f_{{\bf k},{\bf k'}}^{\sigma,\sigma'} &=\>& \sum_\ell
 f_\ell^{\sigma,\sigma'}\cos{\!\big(\ell\theta_{{\bf k},{\bf k'}}\big)} \;,
 \\
 &&\phantom{\sum_\ell}
 f_\ell^{\sigma,\sigma'} \!\equiv 
 \frac{\hbar^2\,\pi}{m^*}\, F_\ell^{\sigma,\sigma'} =
 \frac{m}{m^*}\,\frac{\epsilon_{\scriptscriptstyle\rm F}}{n}\,
 F_\ell^{\sigma,\sigma'} \;.
\end{eqnarray}
At \(P\!=\!0\) the spin--symmetric and --antisymmetric \(\ell\!=\!0,1\) combinations \(F_\ell^{\rm s,a}\equiv \tfrac{1}{2} 
 (F_\ell^{\scriptscriptstyle\uparrow\uparrow}\!\pm 
 F_\ell^{\scriptscriptstyle\uparrow\downarrow})\) 
yield the effective mass \(m^*\), the compressibility \(\kappa\) and the spin--susceptibility. 
These are, in turn, related to the long wavelength behavior of the effective interactions. 
An example is\cite{Iwam91, degiorgi2004strong}
\begin{equation}\begin{array}{rcllc}
 \frac{\kappa^0}{\kappa} &\!=\; 
    \frac{m}{m^*}\,\big(1\!+\! F_0^{\rm s} \big) =
    \frac{1+F_0^{\rm s}}{1+\frac12F_1^{\rm s}}
 \vspace{0.2cm} \;, \\
 1\!-\frac{\kappa^0}{\kappa} &\!=\; 
    \frac{n v(q)}{\epsilon_{\scriptscriptstyle\rm F}} \,G^{\rm s}(q\to 0)\;,
\end{array}\end{equation}
with the \(P=0\) local field corrections \(G^{\rm s}\!\equiv (G_{\uparrow\uparrow}+G_{\uparrow\downarrow})/2\),
and \(\kappa^0\) is the compressibility of free Fermions.
Detailed studies of 
quasiparticles in spin--imbalanced systems are beyond the scope of this work.
It appears interesting, however, to study the slope of our spin dependent
local field corrections.

Following Iwamoto\cite{Iwam91} in defining \(\kappa^0/\kappa\equiv 1\!+\!F^{\rm s}\), 
we investigate the Landau-like parameters
\begin{align}
 F_{\sigma\sigma'} &\equiv\;
 -\textstyle\frac{n}{\epsilon_{\mathrm{F}}}\displaystyle\lim\limits_{q\to 0}\, 
   v(q)\,G_{\sigma\sigma'}(q)
 \nonumber\\\
 \label{eq: Fsspdef} &=\;
 \textstyle\frac{n}{\epsilon_{\mathrm{F}}}\displaystyle\lim\limits_{q\to 0}\, 
  \big[V_{\sigma\sigma'}\!-v(q)\big] \;.
\end{align}
Having ensured the long wave length limit (\ref{eq: Sssp qto0}) of the static 
struc\-ture factors, the effective potentials (\ref{eq: no gentleman}) imply
\begin{subequations}
 \begin{align}
 F_{\uparrow\uparrow} &=  
  \textstyle\frac{\pi^2}{8}\, \left(1\!-\! P\right)^2 - \frac{\pi^2}{4} \;, \\
 F_{\downarrow\downarrow} &= 
  \textstyle\frac{\pi^2}{8}\, \left(1\!+\! P\right)^2 - \frac{\pi^2}{4} \;, \\
 F_{\uparrow\downarrow} &= 
  \textstyle\frac{\pi^2}{8}\, \left( P^2\!-\!1\right) \; .
 \end{align}
\end{subequations}
Consequently, the combination
\(F^\mathrm{a} \equiv F_{\downarrow\downarrow}\!+\!F_{\uparrow\uparrow}-\!2F_{\uparrow\downarrow}\)
vanishes for any \(P\). 
Thus the initial slope of the spin plasmon is not altered compared to 
the RPA result\citep{agarwal2014long}, as also evident from Fig.~\ref{FIG:SPIN_PLASMON_DISPERSION}.
Without additional knowledge, more accurate values of \(F_{\sigma\sigma'}\) must remain 
uncertain\footnote{%
We improved the fit of Ref.\ \protect{\onlinecite{gori2004pair}} by removing an unphysical \(q\ln{q}-\)term; ensuring \(G(q\!\to\!0) \propto\!q\), however, needs further corrections.%
}.

\subsection{Magnetic antiresonance}
\label{ssec: Magnetic anti--resonance}

We conclude our studies by presenting results for dilute systems. 
Knowing the distinct behavior of the various response contributions \(\chi_{\sigma\sigma'}\) allows the identification 
of different \((q,\omega)-\)regions of interest for the imaginary parts of
\(\chi_{nn},\, \chi_{ss}\) and \(\chi_{ns}\), respectively. 
Table \ref{TAB:exciationregimes} shows a comparison of the most prominent cases.

For vanishing \(\uparrow\downarrow\) contributions, density--wave excitations
have the same magnitude as spin--fluctuations; 
contributions to each fluctuating component 
\(\delta n_{\mathbf{q},\sigma} = \chi_{\sigma¸\sigma} V^{\mathrm{ext}}\) 
arise solely from identical spins, the two \(\delta n_\sigma\) react \emph{quasi independently}. If also \(\chi_{\uparrow\uparrow}\) vanishes, the whole excitation arises from the minority spins and the system behaves like a ferromagnetic one. 

\begin{table}[h]
\begin{tabular}{ll}
Condition & Consequence \\\hline
  Im\(\chi_{\uparrow\downarrow}\)=0 & Im\(\chi_{nn}\) = Im\(\chi_{ss}\) 
 \\\hline
  Im\(\chi_{\uparrow\downarrow}\)=Im\(\chi_{\uparrow\uparrow}\)=0 & Im\(\chi_{nn}\) = Im\(\chi_{ss}\) = -Im\(\chi_{ns}\) = Im\(\chi_{\downarrow\downarrow}\) 
 \\\hline
  Im\(\chi_{\uparrow\uparrow}\)=Im\(\chi_{\downarrow\downarrow}\) & 
   \(\!\begin{array}{llc} 
    \mathrm{Im}\chi_{nn} &\!\!=&\! 2\,\mathrm{Im}
     \big(\chi_{\uparrow\uparrow} \!+\!
          \chi_{\uparrow\downarrow} \big) \vspace{0.0cm}\\
    \mathrm{Im}\chi_{ss} &\!\!=&\! 2\,\mathrm{Im}
     \big(\chi_{\uparrow\uparrow} \!-\!
          \chi_{\uparrow\downarrow} \big) \vspace{0.0cm}\\
    \mathrm{Im}\chi_{ns} &\!\!=& 0 \end{array}\)
 \\\hline 
  \(\mathrm{Im}\chi_{\uparrow\downarrow} = \left\{\!\begin{array}{llc} 
    \mathrm{Im}\chi_{\uparrow\uparrow} \vspace{0.0cm}\\
    \mathrm{Im}\chi_{\downarrow\downarrow} \end{array}\right. \)
  &
  \(\!\begin{array}{llc} 
    \mathrm{Im}\chi_{nn} &\!\!=&\! 4\,\mathrm{Im}\chi_{\uparrow\uparrow} 
     \vspace{0.0cm}\\
    \mathrm{Im}\chi_{ss} &\!\!=& 0 \vspace{0.0cm}\\
    \mathrm{Im}\chi_{ns} &\!\!=& 0 \end{array}\)
 \\\hline 
\end{tabular}
\caption{
Specific excitation regimes and their requirements.  Fig.~\ref{FIG: different domains} compares all these loss functions for one characteristic \(q-\)value as function of frequency. 
\label{TAB:exciationregimes}}
\end{table}

In partially spin--polarized systems, at \((q,\omega)-\)values with \(\text{Im}\,\chi_{\uparrow\uparrow}\! = \text{Im}\,\chi_{\downarrow\downarrow}\,\), 
the excitation spectrum is quali\-tatively the same as for the \emph{paramagnetic} case.

\begin{figure}[h!]
\includegraphics[width=0.5\textwidth]{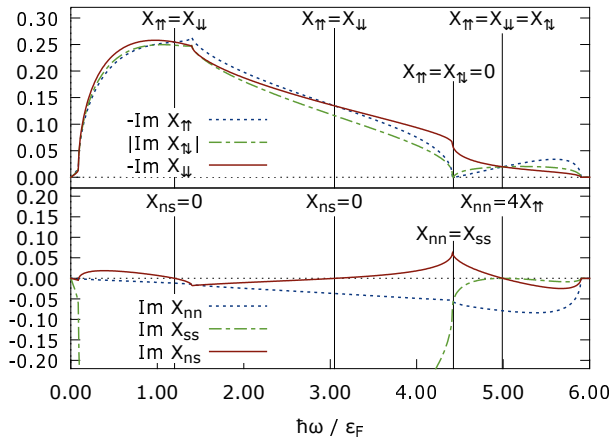}
\caption{
Comparison of the various excitation domains (see Table \ref{TAB:exciationregimes})
for \(r_{\mathrm{S}}\!=\! 20\), \(P\!=\! 0.48\) at \(q \!=\! 1.5\,k_{\mathrm{F}}\).
All response functions are in \(n/\epsilon_{\scriptscriptstyle\mathrm{F}}\). The charge plasmon is outside the shown range, at \(\hbar\omega_{\mathrm{pl}}\!\approx\!6.5\epsilon_{\scriptscriptstyle\mathrm{F}}\).
The vertical lines mark equalities of imaginary parts, the corresponding real parts do not coincide.
}
\label{FIG: different domains}
\end{figure}

Furthermore, a totally new structure emerges in the majority particle--hole band: 
The imaginary part of \(\chi_{ss}\) vanishes exactly along a line
\(\omega_{\scriptscriptstyle\mathrm{mAR}}(q)\) and stays very small in its neighborhood. 
Before discussing this in more detail, we compare the imaginary parts of the various response functions in Fig.\,\ref{FIG: different domains} to identify the excitation regimes given in Table \ref{TAB:exciationregimes}.

First, we observe that all three \textit{partial\/} response functions behave very similar throughout the region \(\hbar\omega \!\lesssim\!  4.2\,\epsilon_{\scriptscriptstyle\mathrm{F}}\), leading, however, to significant differences in the spin summed response functions. 
In this range the imaginary part of the spin--spin response function Im\(\chi_{ss}\) dominates over Im\(\chi_{nn}\) and Im\(\chi_{ns}\). 
Second, we notice that at three non--trivial points the scattering response functions for like spins coincide, Im\(\chi_{\uparrow\uparrow}\) = Im\(\chi_{\downarrow\downarrow\,}\).
Naturally, this is highly sensitive to the effective interactions \(V_{\sigma\sigma'}\); in Fig.~\ref{FIG: different domains} it occurs at
\(\hbar\omega/\epsilon_{\scriptscriptstyle\mathrm{F}} \approx 1.1,\, \gtrsim\!3\), and 5.
At these frequencies Im\(\chi_{ns}\) vanishes, the scattering appears para\-magnetic. 

Third, at \(\hbar\omega /\epsilon_{\scriptscriptstyle\mathrm{F}} \approx 4.5\), both, Im\(\chi_{\uparrow\downarrow}\) and Im\(\chi_{\uparrow\uparrow}\) vanish. 
Here, the whole excitation spectrum is given by the response of the minority spin electrons (red solid line in Fig.~\ref{FIG: different domains}). 
The frequency \(5\,\epsilon_{\scriptscriptstyle\mathrm{F}}/\hbar\) is particularly interesting, as there the imaginary parts of all three partial response functions became equal.  
No magnetic resonance is
possible, \(\mathrm{Im}\chi_{ss} =\!0\!= \mathrm{Im}\chi_{ns}\).
Fig.~\ref{FIG: different domains} also shows the smallness and flatness of
Im\(\chi_{ss}\) in the vicinity of this zero.

\begin{figure}[H]
 \includegraphics[width=0.5\textwidth]{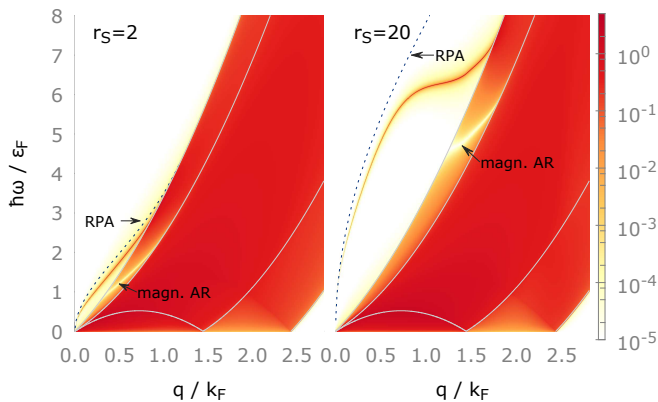}
 \caption{
  GRPA imaginary part of the longitudinal spin density response function 
  \(-\mathrm{Im}\,\chi_{ss}(q,\omega)\) (in units \(\pi\hbar^2/m\) ) for two different 
  densities.
  The grey lines are the characteristic frequencies of the electron--hole continua; 
  the blue dotted line is the RPA charge plasmon.
  The spin polarization is \(P\!=\!0.48\).
 }
 \label{FIG: contourchiss}
\end{figure}

The excitation spectrum for the longitudinal magnetization resulting from \(\mathrm{Im}\,\chi_{ss}\) is shown in Fig.\ \ref{FIG: contourchiss} for moderately high (left) and rather low (right) densities (and, again, \(N_\uparrow\!\approx\!3N_\downarrow\)).
For high \(r_{\scriptscriptstyle\mathrm{S}}\) the charge plasmon develops a flat region at intermediate wave vectors, related to \(S(q)\) there being significantly larger than its RPA counterpart.
This implies the considerably lower \(q^{\mathrm{max}}_\mathrm{pl}\) reported in Table 
\ref{TAB:vphresults}.
The longitudinal spin plasmon is too weak to be visible.
A prominent feature in Fig.~\ref{FIG: contourchiss} is the white valley around \(\omega_{\scriptscriptstyle\mathrm{mAR}}(q)\) on the left side of the continua.
The physical explanation of this intriguing effect is currently not yet fully clarified.
This gap in \(\mathrm{Im}\,\chi_{ss}\) is different from the ``pseudogap'' found in Ref.\,\onlinecite{agarwal2014long} for \(\mathrm{Im}\,\epsilon^{-1}\!\propto\mathrm{Im}\,\chi_{nn}\). 

The antiresonance gap is also present in the bare RPA; the usage of the spin dependent effective potentials of Eq.\ (\ref{eq: no gentleman}), again, shifts \(\omega_{\scriptscriptstyle\mathrm{mAR}}(q)\) towards lower energies.  
We term it ``\textit{magnetic antiresonance\/}'' for the following reason: 
\(\omega_{\scriptscriptstyle\mathrm{mAR}}(q)\) does not describe a collective excitation,  the real parts of \(\chi_{ss}\) and \(\chi_{ns}\) being finite, while the imaginary parts of both response functions vanish. 
Therefore, along this line, neither contributions of spin fluctuations \(\,\delta s\,\) nor of density fluctuations \(\,\delta n\,\) to the double--differential cross section are caused by resonances with a magnetic disturbance \(b^{\mathrm{ext}}\).
(Conversely, \(V^{\mathrm{ext}}\big(q,\omega_{\scriptscriptstyle\mathrm{mAR}}(q)\big)\)
does also not cause spin fluctuations in \(P\!\ne\!0\) systems). The imaginary part of the permeability determines the magnetic loss in 
dispersive media\cite{[{
See, ch. 9 in }][{}]landau1984electrodynamics}.
Although in practical applications transverse precession plays a 
significant role, the suppression of any dissipation channel is highly
desirable. The vanishing of the longitudinal \(\mathrm{Im}\,\chi_{ss}\) 
 around \(\omega_{\scriptscriptstyle\mathrm{mAR}}(q)\) 
is therefore very promising.
As known for Lorentzian fits of experiments, the maximum
in the imaginary part occurs at the resonance frequency, where the real part
vanishes. This further supports the name 'magnetic antiresonance'.

From Fig.~\ref{FIG: contourchiss} it is obvious that the magnetic antiresonance is observed in the particle--hole band of the majority spins.  This facilitates the calculation of its dispersion relation.
Introducing the dimensionless potential
\begin{equation}
  \label{eq: VmAR}
  V_{\!_{\downarrow+\!}} \equiv\>
  \textstyle\frac{n}{\epsilon_{\scriptscriptstyle\rm F}}\,\displaystyle
  \big(V_{\downarrow\downarrow}(q) + V_{\uparrow\downarrow}(q) \big) \;,
\end{equation}
we obtain
\begin{equation}
 \label{eq: mAR dispersion relation}
 \begin{array}{lll}\displaystyle
  \frac{\hbar^2\omega_{\scriptscriptstyle\mathrm{mAR}}^2(q)}
       {\epsilon_{\scriptscriptstyle\rm F\downarrow}^2 } &\!=&\!\displaystyle
  \frac{q^2/k_{\scriptscriptstyle\rm F\downarrow}^2}{V_{\!_{\downarrow+\!}}(q)}\,
  \Big(2+V_{\!_{\downarrow+\!}}(q) \Big)^{\!2}
  \vspace{0.2cm}\\ &&\displaystyle\qquad
  \times \bigg(\frac{V_{\!_{\downarrow+\!}}(q)}{1+V_{\!_{\downarrow+\!}}(q)} \,+\,
      \frac{q^2/k_{\scriptscriptstyle\rm F\downarrow}^2}{V_{\!_{\downarrow+\!}}(q)} \bigg) \;.
 \end{array}
\end{equation}
It is straightforward to show that the upper minority spin band is tangentially
hit at 
\begin{equation}
  \frac{2q^{\mathrm{c}\downarrow}_{_\mathrm{mAR}}}{k_{\scriptscriptstyle\rm F\uparrow}} \>=\>
\frac{V_{\!_{\downarrow+\!}}^2}{4\big(1+V_{\!_{\downarrow+\!}}\big)} \;,
\end{equation}
and the upper majority spin band is cut at
\begin{equation}
  \frac{2q^{\mathrm{c}\uparrow}_{_\mathrm{mAR}}}{k_{\scriptscriptstyle\rm F\uparrow}} \>=\>
  \frac{V_{\!_{\downarrow+\!}}^2 + \sqrt{\frac{2P}{1-P}} 
   \big(V_{\!_{\downarrow+\!}}^2+\!2V_{\!_{\downarrow+\!}}\big) }{1+V_{\!_{\downarrow+\!}}} \;.
\end{equation}

Both, the spin--spin as well as the density--spin response function take a very simple form along \(\omega_{\scriptscriptstyle\mathrm{mAR}}(q)\,\),
\begin{equation}
 \chi_{\!_{\scriptstyle ss\atop (ns)\!}}\big(q, \omega_{\scriptscriptstyle\mathrm{mAR}}(q)\big)
 \>=\> {\textstyle \genfrac{}{}{0pt}{}{+}{(-)} } \frac{1}{V_{\uparrow\downarrow}(q)} \;.
 \label{eq: ChissmaR}
\end{equation}
The mAR features of the bare RPA are obtained by replacing \(V_{\!_{\downarrow+\!}}(q)\) with \(2n v(q)/\epsilon_{\scriptscriptstyle\mathrm{F}}\) in Eqs.(\ref{eq: mAR dispersion relation}-\ref{eq: ChissmaR}).

In a realistic scattering experiment spin channels have to be taken into account in the double--differential cross section, well explained by Perez~\cite{perez2009spin}. 
How exactly the magnetic antiresonance contributes~\cite{kreil2014spin} to this cross section depends on the size of the optical matrix elements.
\section{Conclusion}
\label{sec: Conclusion}
In summary, we have shown that exchange- and correlation effects substantially alter the response functions of partially spin--polarized electron layers compared to the bare RPA.
In particular, the spin plasmon is shifted downwards and its stability region is severely decreased.
For the charge plasmon our results are in good agreement with those obtained from literature--based local field corrections \cite{davoudi2001analytical}. 
Finally, we predict a new structure in the double--differential cross section, characterized by a zero in the imaginary part of the spin--spin response function.  Certainly, this interesting region and the implications of this effect deserve further research.
\section*{Acknowledgments}
We thank Paola Gori-Giorgi for the Quantum Monte Carlo data of the
spin--resolved pair--distribution functions with helpful comments 
and Martin Panholzer for valuable discussions.
\bibliographystyle{apsrev4-1}
\bibliography{references}
\end{document}